\documentclass[twocolumn,prl,showpacs,preprintnumbers,amsmath,amssymb]{revtex4}

\usepackage{graphicx}
\usepackage{dcolumn}
\usepackage{bm}

\def\a{\alpha}		 		   	
         \def\s{\sigma}       	   	
\def\l{\lambda}	         \def\t{\tau}					
\def\D{\Delta}	         \def\W{\Omega}		   	
          \def\j{{\bf j}}				
	 
\def\AJ{A_{\rm J}}       \def\AN{A_{\rm N}}	   
\def\RN{{ R}_{\rm N}}    \def\RF{{ R}_{\rm F}}			
       
\def\lN{\l_{\rm N}}	 \def\lF{\l_{\rm F}}	   \def\lS{\l_{\rm S}}
\def\sN{\s_{\rm N}}              	 
     
\def\muF{\mu_{\rm F}}    \def\muFi{\mu_{{\rm F}i}}
\def\muN{\mu_{\rm N}}    \def\dmuN{\delta\mu_{\rm N}} 
\def\rN{\rho_{\rm N}}    \def\rF{\rho_{\rm F}}     \def\sN{\s_{\rm N}}	
\def\dN{d_{\rm N}}       \def\dF{d_{\rm F}}     
\def\wN{w_{\rm N}}       \def\wF{w_{\rm F}}     
\def\us{\uparrow}         \def\ds{\downarrow}	
\def\dmu{\delta\mu}	 \def\kF{k_{\rm F}}	   
\def\[{\left[}           \def\]{\right]}	   
\def\({\left(}           \def\){\right)}	   
\def\<{\langle}          \def\>{\rangle}	   
\def\kT{k_{\rm B}T}	          \def\tsf{\tau_{\rm sf}}
\def\timp{\tau_{\rm imp}}

\def\pF{p_{\rm F}}	 \def\PT{P_{\rm T}}        
\def\dt{\partial t}	 \def\d{\partial}         

\begin{document}


\title{Spin Injection and Nonlocal Spin Transport in Magnetic Nanostructures}

\author{S. Takahashi and S. Maekawa}

\affiliation{
Institute for Materials Research, Tohoku University, Sendai 980-8577, Japan,
\\ and CREST, Japan Science and Technology Corporation, Kawaguchi 332-0012, 
Japan}

\date{September 1, 2005}

\begin{abstract}
We theoretically study the nonlocal spin transport in a device 
consisting of a nonmagnetic metal (N) and ferromagnetic injector
(F1) and detector (F2) electrodes connected to N.
We solve the spin-dependent transport equations in a device with 
arbitrary interface resistance from a metallic-contact to tunneling 
regime, and obtain the conditions for efficient spin injection, 
accumulation, and transport in the device.  
In a device containing 
a superconductor (F1/S/F2), the effect of superconductivity on 
the spin transport is investigated.  The spin-current induced 
spin Hall effect in nonmagnetic metals is also discussed.
\end{abstract}



\maketitle


\bigskip
\bigskip
\noindent
{\bf \normalsize 1. Introduction}
\bigskip

There has been considerable interest in spin transport in magnetic 
nanostructures, because of their potential applications as spin-electronic 
devices \cite{book}.  
The spin polarized electrons injected from a ferromagnet (F) into a 
nonmagnetic material (N) such as a normal metal, semiconductor, and 
superconductor create a nonequilibrium spin accumulation in N.
The efficient spin injection, accumulation, and transport are central 
issues for utilizing the spin degree of freedom as
in spin-electronic devices.  
It has been demonstrated that the injected spins penetrate 
into N over the spin-diffusion length ($\lN$) of the order of 
1{\,}$\mu$m using spin injection and detection technique in F1/N/F2 
trilayer structures (F1 is an injector and F2 a detector) \cite{johnson}.  
Recently, several groups have succeeded in observing spin 
accumulation by the {\it nonlocal} spin injection and detection technique
  \cite{jedemaCu,jedemaAl,kimuraAPL,urech,ji,garzonPRL,miuraJMMM}.  

In this paper, we study the spin accumulation and spin current, and
their detection in the nonlocal geometry of a F1/N/F2 nanostructure.   
We solve the diffusive transport equations for the electrochemical 
potential (ECP) for up and down spins in the structure of arbitrary 
interface resistances ranging from a metallic-contact to a tunneling regime,
and examine the optimal conditions for spin accumulation and spin current.  
Efficient spin injection and detection are achieved when a tunnel 
barrier is inserted at the interface, whereas a large spin-current 
injection from N into F2 is realized when N is in metallic 
contact with F2, because F2 plays the role of strong spin absorber. 
In a tunnel device containing a superconductor (F1/S/F2), the effect 
of superconductivity on the spin transport is discussed. 
The spin-current induced anomalous Hall effect is also discussed.

\bigskip
\bigskip
\noindent
{\bf \normalsize 2. Spin injection and accumulation}
\bigskip

\begin{figure}[b]					
  \begin{center}
    \includegraphics[width=0.88\columnwidth]{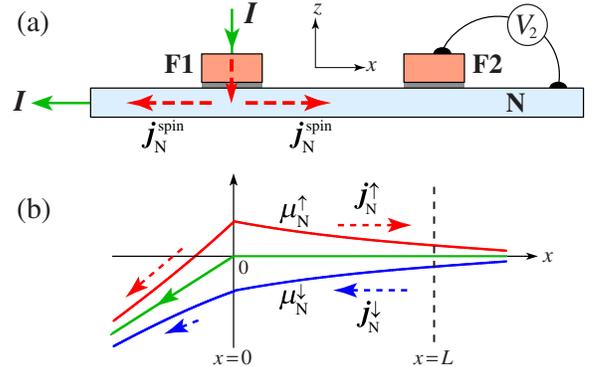}
  \end{center}
  \caption{						
(a) Spin injection and detection device (side view).
The current $I$ is applied from F1 to the left
side of N.  The spin accumulation at $x=L$ is detected
by measuring voltage $V_2$ between F2 and N.					
(b) Spatial variation of the electrochemical potential
(ECP) for up and down spin electrons in N.
   }   \label{fig1}					
\end{figure}

We consider a spin injection and detection device consisting of 
a nonmagnetic metal N connected to ferromagnetic injector F1
and detector F2 as shown in Fig.~\ref{fig1}.
The F1 and F2 are the same ferromagnets of width $\wF$
and thickness $\dF$ and are separated by distance $L$,
and N of of width $\wN$ and thickness $\dN$.
The magnetizations of F1 and F2 are aligned either parallel or
antiparallel.

In the diffusive spin transport, the current $\j^\s$ for spin channel
$\s$ in the electrodes is driven by the gradient of ECP ($\mu^\s$)
according to
    $\j^\s = -{(1/e\rho^\s) } \nabla \mu^\s$,	
where $\rho^\s$ is the resistivity.
The continuity equations for the charge and spin curents 
in a steady state yield
    \cite{johnson,valet,fertlee,hershfield,takahashiPRB}
\begin{eqnarray}				
  & & \nabla^2 \( \mu^\us/\rho^\us + \mu^\ds/\rho^\ds \) = 0, 
   \label{eq:ddr1}	\\		
  & & \nabla^2 \(\mu^\us - \mu^\ds\)		
    = \l^{-2} \(\mu^\us - \mu^\ds\),	
   \label{eq:ddr2}				
\end{eqnarray}					
where $\l$ is the spin-diffusion length and takes
$\lN$ in N and $\lF$ in F.
We note that $\lN$ ($\l_{\rm Cu} \sim 1\mu$m \cite{jedemaCu},
$\l_{\rm Al} \sim 1\mu$m 
   \cite{johnson,jedemaAl})
is much larger than $\lF$
($\l_{\rm Py} \sim 5${\,}nm, $\l_{\rm CoFe} \sim 12${\,}nm,
 $\l_{\rm Co} \sim 50${\,}nm) \cite{bass}.

We employ a simple model for the interfacial current across the junctions
\cite{valet}. 
Due to the spin-dependent interface resistance $R^\s_i$ ($i=1,2$),
the ECP is discontinuous at the interface, and
the current $I_i^{\s}$ across the interface ($z=0$) is given by
    $I_i^{\s} = (1/eR_i^\s) \( \muF^\s|_{z=0^+} - \muN^\s|_{z=0^-} \)$,
where the current distribution is assumed to be uniform over
the contact area \cite{ichimura,hamrle}.
In a transparent contact (tunnel junction) 
the discontinuous drop in ECP is much smaller (larger) than the spin splitting
of ECP.
The interfacial charge and spin currents are
$I_i=I_{i}^{\us}+I_{i}^{\ds}$ and $I_i^{\rm spin}=I_{i}^{\us}-I_{i}^{\ds}$.

When the bias current $I$ flows from F1 to the left side of N ($I_1=I$),
there is no charge current on the right side ($I_2=0$).
The solution for Eqs.~(\ref{eq:ddr1}) and (\ref{eq:ddr2}) takes the form
   $\muN^\s(x) = {\bar \muN} + \s \dmuN$	
with the average 
${\bar \muN} = -{(eI\rN/\AN)}x$ for $x<0$ and ${\bar \muN} = 0$ for $x>0$,
and the splitting $\dmuN = a_1 e^{-|{x}|/\lN} - a_2 e^{-|x-{L}|/\lN}$,
where the $a_1$-term represents the spin accumulation due to 
spin injection at $x=0$, while the $a_2$-term the decrease of 
spin accumulation due to the contact of F2.  Note that the pure spin 
current
 $I_{\rm N}^{\rm spin}=I_{\rm N}^{\us}-I_{\rm N}^{\ds}$
flows in the region of $x>0$.

In the F1 and F2 electrodes, the solution takes the form
   $\muFi^\s(z) = {\bar \muFi} + \s b_i\(\rF^\s/{\rF}\) e^{-z/\lF}$,
with ${\bar \mu}_{\rm F1} = -{(eI\rF/\AJ)}z + eV_1$ in F1
and  ${\bar \mu}_{F2} = eV_2$ in F2, 
where $V_1$ and $V_2$ are the voltage drops across junctions 1 and 2, 
and $\AJ=\wN\wF$ is the contact area of the junctions.  

Using the matching condition for the spin current at the interfaces, 
we can determine the constants $a_i$, $b_i$, and $V_i$.
The spin-dependent voltages detected by F2 are $V^{\rm P}_2$ and $V^{\rm AP}_2$
for the parallel (P) and antiparallel (AP) alignment of magnetizations.  
The spin accumulation signal detected by F2,
 $R_s=(V^{\rm P}_2-V^{\rm AP}_2)/I$, is given by \cite{takahashiPRB}
\begin{widetext}
  \begin{eqnarray}					
    {R_s}  =  {4 \RN}
    \frac{\displaystyle
    \left({P_1 \over 1-P_1^2} {R_1 \over \RN} +
    {\pF \over 1-\pF^2} {\RF \over \RN}  \right) 
    \left({P_2 \over 1-P_2^2} {R_2 \over \RN} +
    {\pF \over 1-\pF^2} {\RF \over \RN}  \right)
    e^{-{L/\lN}}
    }{\displaystyle
    \left( 1 + {2 \over 1-P_1^2} {R_1 \over \RN}
    +{2 \over  1-\pF^2} {\RF \over \RN} \right)
    \left( 1 + {2 \over 1-P_2^2} {R_2 \over \RN}
    +{2 \over  1-\pF^2} {\RF \over \RN} \right)
     - e^{-{2L/\lN}} ,
     }
     \label{eq:V2}			
  \end{eqnarray}					
\end{widetext}
where $\RN = \rN\lN/\AN$ and $\RF = \rF\lF/\AJ$ are the
{\it spin-accumulation resistances} of the N and F electrodes,
$\AN=\wN\dN$ is the cross-sectional area of N,
$R_i= R_i^\uparrow+R_i^\downarrow$ is the interface resistance 
of junction $i$, 
$P_i = {| R_i^\uparrow-R_i^\downarrow |/R_i}$
is the interfacial current spin-polarization,
and $\pF = {| \rF^\uparrow-\rF^\downarrow |/\rF}$
is the spin-polarization of F.
In metallic contact junctions, the spin polarizations, $P_i$ and $\pF$, 
range around 40--70\%   
from GMR experiments \cite{bass} and point-contact Andreev-reflection
experiments \cite{soulen}, 
whereas in tunnel junctions, $P_i$ r anges around 30--55\% from 
superconducting tunneling spectroscopy experiments with alumina tunnel
barriers \cite{meservey,moodera-mathon,monsma-parkin},
and $\sim 85\, \%$ in MgO barriers
 \cite{parkin-MgO,yuasa-MgO}.

The spin accumulation signal $R_s$ strongly depends on whether
each junction is either a metallic contact or a tunnel junction.
By noting that there is large disparity between $\RN$ and $\RF$
($\RF/\RN$ $\sim 0.01$  for Cu and Py \cite{jedemaCu}),
we have the following limiting cases. 
When both junctions are transparent contact ($R_1,R_2 \ll \RF$), 
we have \cite{jedemaCu,fertlee,hershfield}
  \begin{eqnarray}			
    R_s/\RN =  {2\pF^2 \over  (1-\pF^2)^2} 	
    {\(\RF\over \RN\)^2} \sinh^{-1}(L/\lN) .	
     \label{eq:Rs-mm}			
  \end{eqnarray}			
When junction 1 is a tunnel junction and junction 2 is a transparent 
contact ({\it e.g.}, $R_2 \ll \RF \ll \RN \ll R_1$), we have
\cite{takahashiPRB}
  \begin{eqnarray}			
    R_s/\RN =  {2\pF P_1 \over  (1-\pF^2)} {\(\RF\over \RN\)}
    {e^{-{L/\lN}} } .	
     \label{eq:Rs-tm}			
  \end{eqnarray}			
When both junctions are tunnel junctions ($R_1,R_2 \gg \RN$),
we have \cite{johnson,jedemaAl}
  \begin{eqnarray}		
    R_s/\RN = P_1P_2 e^{-{L/\lN}},
     \label{eq:Rs-tt}		
  \end{eqnarray}		
where $\PT=P_1=P_2$.
Note that $R_s$ in the above limiting cases is independent of $R_i$.

\begin{figure}[b]					
  \begin{center}
    \includegraphics[width=0.88\columnwidth]{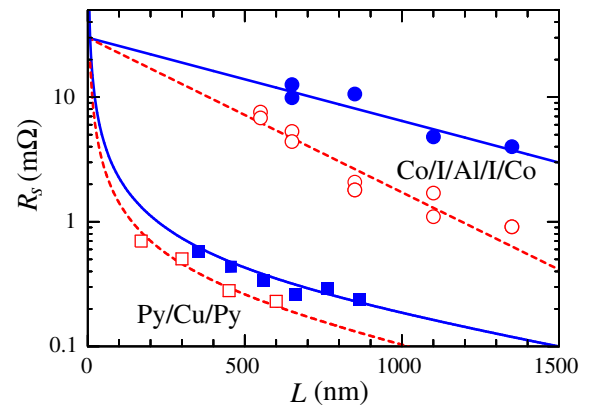}
  \end{center}
  \caption{					
Spin accumulation signal $R_s$ as a function of distance $L$ between
the ferromagnetic electrodes in tunnel devices:
   ($\bullet,\circ$) Co/I/Al/I/Co \protect\cite{jedemaAl},
and in metallic-contact devices: 
   ($\square,\blacksquare$) Py/Cu/Py \protect\cite{garzonPhD,kimuraJMSJ},
where ($\bullet,\blacksquare$) are the data 
at 4.2K and ($\circ, \square$) at room temperature.
  }						
  \label{fig2}						
\end{figure}					

We compare our theoretical result to experimental data measured
by several groups.  Figure~\ref{fig2} shows the theoretical curves
and the experimental data of $R_s$ as a function of $L$.
The solid curves are the values in a tunnel device,
and the dashed curves are those in a metallic-contact device.
We see that $R_s$ in a metallic contact device is smaller 
by one order of magnitude than $R_s$ in a tunnel device, 
because of the resistance mismatch ($\RF/\RN\ll 1$).
Fitting Eq.~(\ref{eq:Rs-tt}) to the experimental data of 
Co/I/Al/I/Co (I{\,}={\,}Al$_2$O$_3$) in Ref.{\,}\cite{jedemaAl} yields 
$\lN =650$~nm (4.2 K), $\lN =350${\,}nm (293{\,}K), $P_1 = 0.1$,
and $\RN=3{\,}\W$.
Fitting Eq.{\,}(\ref{eq:Rs-mm}) to the 
data of Py/Cu/Py in Ref.{\,}\cite{garzonPhD} at 4.2{\,}K yields 
$\lN =920${\,}nm, $\RN=5{\,}\W$,
$[\pF/(1-\pF^2)](\RF/\RN) = 5 \times 10^{-3}$, and 
fitting to the data in Ref.{\,} \cite{kimuraJMSJ} at 293{\,}K yields 
$\lN =700${\,}nm,
$\RN=1.75{\,}\W$, and $[\pF/(1-\pF^2)](\RF/\RN) = 8 \times 10^{-3}$.

The spin splitting in N in the tunneling case is
  \begin{eqnarray}			
   2\dmuN(x)={P_1} e\RN I e^{-|x|/\lN}.
  \label{eq:dmuN}
  \end{eqnarray}			
In the case of Co/I/Al/I/Co, $\dmuN(0) \sim 15${\,}$\mu$V for 
$P_1 \sim 0.1$, $\RN=3{\,}\W$, 
and $I=100${\,}$\mu$A
  \cite{jedemaAl},
which is much smaller than the superconducting gap 
$\D \sim 200$$\mu$eV of an Al film.

\bigskip
\bigskip
\noindent
{\bf 3. Nonlocal spin injection and manipulation}
\bigskip

We next study how the spin-current flow in the structure is affected by 
the interface condition, especially, the spin current through the N/F2 
interface, because of the interest in spin-current induced magnetization 
switching \cite{slonczewski}.

The spin current injected nonlocally across the N/F2 interface is given by
 \cite{takahashiPRB}
\begin{widetext}
  \begin{eqnarray}					
    I^{\rm spin}_{\rm N/F2}  =  2I 
    \frac{\displaystyle
     \left({P_1 \over 1-P_1^2} {R_1 \over \RN} +
    {\pF \over 1-\pF^2}  {\RF \over \RN} \right)
    e^{-{L/\lN}}
    }{\displaystyle
     \left( 1 + {2 \over 1-P_1^2} {R_1 \over \RN}
    +{2 \over 1-\pF^2} {\RF \over \RN} \right)
     \left( 1 + {2 \over 1-P_2^2} {R_2 \over \RN}
    +{2 \over  1-\pF^2} {\RF \over \RN} \right)
    - e^{-{2L/\lN}} 
    } .
     \label{eq:IF2}			
  \end{eqnarray}					
\end{widetext}
A large spin-current injection occurs when junction 2 is a 
metallic contact ($R_2 \ll \RN$) and junction 1 is a tunnel junction 
($R_1 \gg \RN$), yielding
  \begin{eqnarray}					
    I^{\rm spin}_{\rm N/F2} \approx  P_1  I e^{-{L/\lN}} ,
     \label{eq:IF2-tm}			
  \end{eqnarray}					
for F2 with very short $\lF$. 
The spin current flowing in N on the left side of F2 is
$I^{\rm spin}_{\rm N}=P_1  I e^{-{x/\lN}}$, which is two times larger 
than that in the absence of F2, while on the right side left
$I^{\rm spin}_{\rm N} \approx 0$. 
This indicates that F2 like Py and CoFe works as a strong absorber
(sink) for spin current, providing a method for magnetization reversal
in nonlocal devices with reduced dimensions of F2 island
\cite{kimuraCondmat}.

\bigskip
\noindent
{\bf \normalsize 4. Spin injection into superconductors}
\bigskip

The spin transport in a device containing a superconductor (S) 
such as Co/I/Al/I/Co is of great interest,
because $R_s$ is strongly influenced by opening the superconducting gap.
In such tunneling device, the spin signal would be strongly affected
by opening the superconducting gap $\D$.

We first show that the spin diffusion length in the superconducting 
state is the same as that in the normal state \cite{yamashitaPRB,morten}.
This is intuitively understood as follows. 
Since the dispersion curve of the quasiparticle (QP) excitation energy is
given by $E_k=\sqrt{\xi^2_k+\D^2}$ with one-electron energy $\xi_k$
\cite{tinkham},
the QP's velocity 
$\tilde{v}_k=(1/\hbar)(\d E_k/\d k) = (|\xi_k|/E_k) v_{k}$
is slower by the factor $|\xi_k|/E_k$ compared with the normal-state
velocity $v_k (\approx v_F )$.
By contrast, the impurity scattering time \cite{bardeen} 
$\tilde{\tau}=(E_k/|\xi_k|)\tau$
is longer by the inverse of the factor.
Then, the spin-diffusion length  in S,
  $
      \lS=(\tilde{D}\tilde{\t}_{sf})^{1/2}
  $
with
$\tilde{D}=\frac{1}{3} \tilde{v}^2_{k}\tilde{\t}_{\rm tr}=(|\xi_k|/E_k)D$
turns out to be the same as $\lN$, owing to the cancellation
of the factor $|\xi_k|/E_k$.

\begin{figure}[t]					
  \begin{center}
    \includegraphics[width=0.88\columnwidth]{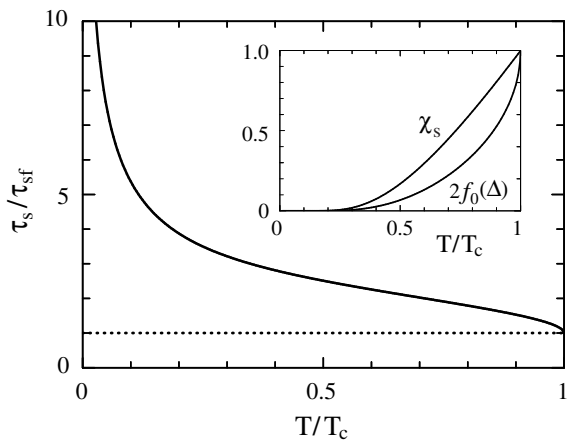}
  \end{center}
  \caption{						
Temperature dependence of the spin relaxation time 
$\tau_s$ in the superconducting state.  The inset shows 
$\chi_s$ and $2f_0(\D)$ vs. $T$.	
   }							
  \label{fig3}					
\end{figure}

The spin accumulation in S is determined by balancing the spin 
injection rate with the spin-relaxation rate:
  \begin{eqnarray}                          	
   I_1^{\rm spin}- I_2^{\rm spin}+ e\left( {\d S}/{\dt} \right)_{\rm sf}=0,  
     \label{eq:Is-dSdt} 	                        
  \end{eqnarray}                            	
where $S$ is the total spins in S, and $I_1^{\rm spin}$ and 
$I_2^{\rm spin}$ are the rates of incoming and outgoing spin currents 
through junction 1 and 2, respectively.  At low temperatures the spin 
relaxation is dominated by spin-flip scattering via the spin-orbit 
interaction $V_{\rm so}$ at nonmagnetic impurities or grain boundaries.
The scattering matrix elements of $V_{\rm so}$ over QP states 
$|{\bf k}\s\rangle$ with momentum ${\bf k}$ and spin $\s$ has the form:
$  \langle{\bf k}'\s'|V_{\rm so}|{\bf k}\s\rangle = 
  i \eta_{\rm so} \left(u_{k'}u_{k} -v_{k'}v_{k}\right)
  [ {\vec\sigma}_{\s'\s}\cdot ({{\bf k}}\times{{\bf k}'})/\kF^2 ] V_{\rm imp}
$,
where $\eta_{\rm so}$ is the spin-orbit coupling parameter, $V_{\rm imp}$ 
is the impurity potential, ${\s}$ is the Pauli spin matrix, and
  $u_k^2 = 1-v_k^2 = \frac{1}{2}\left( 1+{\xi_k/E_k}\right)$ are the
coherent factors \cite{tinkham}.
Using the golden rule for spin-flip scattering processes,
we obtain the spin-relaxation rate in the form
\cite{takahashiJMMM,yafet}
  \begin{eqnarray}   	                 	
   \left( {\d S}/{\dt} \right)_{\rm sf} 			
    = - {S}/{\tsf(T)},
     \label{eq:dSdt} 	                        
  \end{eqnarray}                            	
where $S=\chi_s(T)S_N$ with $S_N$ the normal-state value and 
$\chi_s(T)$ the QP spin-susceptibility called the Yosida function 
\cite{yosida}, and
  \begin{eqnarray}                      
    \tau_{s}(T) = \left[ {\chi_s(T)}/{2f_0(\D)} \right] \tsf,	
     \label{eq:tS-Yafet}                
  \end{eqnarray}                        
where $\tsf$ is the spin-flip scattering time in the normal state.
Equation~(\ref{eq:tS-Yafet}) was derived earlier by Yafet \cite{yafet} 
who studied the electron-spin resonance (ESR) in the superconducting state.
Figure~\ref{fig3} shows the temperature dependence of $\tau_{s}/\tsf$.
In the superconducting state below the superconducting 
critical temperature $T_c$, 
$\tau_{s}$ becomes longer with decreasing $T$ according to
$\tau_{s} \simeq (\pi\D/2\kT)^{1/2}\tsf$ at low temperatures.

Since the spin diffusion length in the superconducting state is the 
same as that in the normal state, the ECP shift in S is
$\delta\mu_{\rm S}=\left(\tilde{a}_1 e^{-|{x}|/\lN} 
                  - \tilde{a}_2 e^{-|x-{L}|/\lN}\right)$,
where $\tilde{a}_i$ is calculated as follows.  
In the tunnel device, the tunnel spin currents are $I_1^{\rm spin}=P_1I$ 
and $I_2^{\rm spin}\approx 0$, so that
Eqs~(\ref{eq:Is-dSdt}) and (\ref{eq:dSdt}) give the coefficients 
$\tilde{a}_1=P_1\RN eI/[2 f_0(\D)]$ and $\tilde{a}_2 \approx 0$,
leading to the spin splitting of ECP in the superconducting state
\cite{takahashiPRB}
  \begin{eqnarray}			
    \delta\mu_{\rm S}(x)=  \frac{1}{2} P_1 
      \frac{\RN e I}{2f_0(\D)}  {e^{-{|x|/\lN}} },
     \label{eq:H-dmuN}			
  \end{eqnarray}			
indicating that the splitting in ECP is enhanced by the
factor $1/[2f_0(\D)]$ compared with the normal-state value
(see Eq.{\,}7).
The detected voltage $V_2$ by F2 at distance $L$ is given by
$V_2=\pm P_2 \delta\mu_{\rm S}(L)$  for the P ($+$) and AP ($-$) 
alignments.  Therefore, the spin signal $R_s$
in the superconducting state becomes \cite{takahashiPRB}
  \begin{eqnarray}			
    R_s = P_1P_2  {\RN} {e^{-{L/\lN}}} /[{2f_0(\D)}].	
     \label{eq:H-Rs}			
  \end{eqnarray}			
The above result is also obtained by the replacement
$\rN \rightarrow \rN/[2f_0(\D)]$ in the normal-state result of
Eq.~(\ref{eq:Rs-tt}), which results from the fact that the QP
carrier density decreases in proportion to $2f_0(\D)$, and 
superconductors become a low carrier system for spin transport.
The rapid increase in $R_s$ below $T_c$ reflects the strong reduction
of the carrier population.  However, when the splitting
$\dmu_{\rm S} \sim \frac{1}{2}e P_1 \RN I/[2f(\D)]$ at $x=0$ becomes 
comparable to or larger than $\D$, the superconductivity is suppressed 
or destroyed by pair breaking due to the spin splitting
\cite{takahashiPRL82,takahashiJAP,chenPRL88,johanssonJAP,wangJAP,daibou}.
This prediction can be tested by measuring $R_s$ in Co/I/Al/I/Co 
or Py/I/Al/I/Py in the superconducting state.

\bigskip
\bigskip
\noindent
{\bf \normalsize 5. Spin-current induced spin Hall effect}
\bigskip

The basic mechanism for the spin Hall effect (SHE) is the spin-orbit 
interaction in N, which causes a spin-asymmetry
in the scattering of conduction electrons by impurities;
up-spin electrons are preferentially scattered in one direction
and down-spin electrons in the opposite direction.  
Spin injection techniques makes it possible
to cause SHE in {\it nonmagnetic} conductors.
When spin-polarized electrons are injected from a ferromagnet (F) to
a nonmagnetic electrode (N), these electrons moving in N are deflected
by the spin-orbit interaction to induce the Hall current in
the transverse direction and accumulate charge on the sides of N
   \cite{hirsch,zhang,takahashiPRL88}.

\begin{figure}[b]						
  \begin{center}
    \includegraphics[width=0.80\columnwidth]{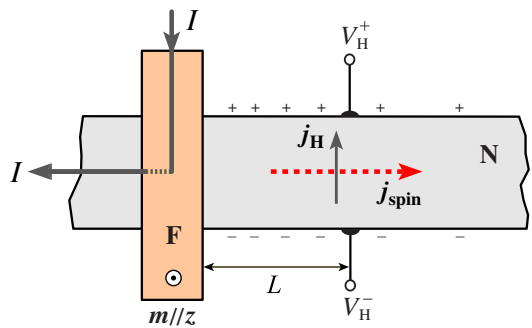}
  \end{center}
  \caption{  
Spin injection Hall device (top view). The magnetic moment 
of F is aligned perpendicular to the plane.  The anomalous
Hall voltage $V_{\rm H}=V^+_{\rm H}-V^-_{\rm H}$ is induced 
in the transverse direction
by injection of	spin-polarized current.					
   }  
\label{fig4}							
\end{figure}						

We consider a spin-injection Hall device shown in Fig.~\ref{fig4}.
The magnetization of F electrode points to the $z$ direction.
Using the Boltzmann transport equation which incorporates the
asymmetric scattering by nonmagnetic impurities,
we obtain the total charge current in N  \cite{takahashiPRL88}
  \begin{eqnarray}					
   {\bf j}_{\rm tot} 
       = \a_{\rm H} \left[\hat{\bf z} \times {\bf j}_{\rm spin} \right]
       + \sN {\bf E},
   \label{eq:jtot} 				
  \end{eqnarray}					
where the first term is the Hall current ${\bf j}_{\rm H}$ induced by 
the spin current, the second term is the Ohmic current, $ {\bf E}$ is 
the electric field induced by surface charge, and 
$\a_{\rm H} \sim \eta_{\rm so}N(0)V_{\rm imp}$ (skew scattering).
In the open circuit condition in the transverse direction,
the $y$ component of ${\bf j}_{\rm tot}$ vanishes, so that
the nonlocal Hall resistance $R_{\rm H}=V_{\rm H}/I$ becomes
  \begin{eqnarray}		
    R_{\rm H} = \frac{1}{2} \(P_1 \a_{\rm H}\rN/\dN\) e^{-L/\lN},	
     \label{eq:VH}		
  \end{eqnarray}		
in the tunneling case.
Recently, SHE induced by the spin-current have been measured
in a Py/Cu structure using the spin injection technique \cite{kimuraJMMM,valenzuela,kimuraSHE}.

\begin{table}[t]
\begin{center}
  \caption{Spin-orbit coupling parameter of Cu and Al. }
  \label{table1}
{
\begin{tabular}{ccccccccc}
\hline
     & \ \ \ \ & $\lN$ (nm) & \ \ \ \ & $\rN\ (\mu\W$cm) & \ \ \ \ & $\timp/\tsf$  & \ \ \ \ \ & $\eta_{\rm so}$ \\
\hline
\ Cu  && 1000$^{\rm a}$ \ \ &&  1.43$^{\rm a}$  &&  0.70 $\times 10^{-3}$  && 0.040 \\
\ Cu  &&  546$^{\rm b}$ \ \ &&  3.44$^{\rm b}$  &&  0.41 $\times 10^{-3}$  && 0.030 \\
\ Al  &&  650$^{\rm c}$ \ \ &&  5.90$^{\rm c}$  &&  0.36 $\times 10^{-4}$  && 0.009 \\
\ Al  &&  705$^{\rm d}$ \ \ &&  5.88$^{\rm d}$  &&  0.30 $\times 10^{-4}$  && 0.008 \\
\ Ag  &&  195$^{\rm e}$ \ \ &&  3.50$^{\rm e}$  &&  0.50 $\times 10^{-2}$  && 0.110 \\
\hline
\end{tabular}
}
\medskip
\centerline{\small
    $^a$Ref.{\,}\cite{jedemaCu},
    $^b$Ref.{\,}\cite{garzonPRL},
    $^c$Ref.{\,}\cite{jedemaAl},
    $^d$Ref.{\,}\cite{valenzuela},
    $^e$Ref.{\,}\cite{godfrey}.}
\end{center}
\end{table}

It is noteworthy that the product $\rN \lN$ is related to
the spin-orbit coupling parameter $\eta_{\rm so}$ as
\cite{chapter8}
  \begin{eqnarray}                          	
    \rN \lN = \frac{ \sqrt{3}\pi}{2} \frac{R_{\rm K}}{\kF^2} 
                \sqrt{\frac{\tsf}{\timp} }
            = \frac{3\sqrt{3}\pi}{4} \frac{R_{\rm K}}{\kF^2} 
              \frac{1}{\eta_{\rm so} } ,
     \label{eq:lso-rholso} 	               
  \end{eqnarray}                           	
where $R_{\rm K}=h/e^2 \sim 25.8{\,}$k$\W$ is the quantum resistance.
The formula (\ref{eq:lso-rholso}) provides a method for obtaining
information for spin-orbit scattering in nonmagnetic metals. 
Using the experimental data of $\rN$ and $\lN$ and the Fermi momentum
$\kF$ \cite{mermin} in Eq.~(\ref{eq:lso-rholso}), we obtain 
the value of the spin-orbit coupling parameter $\eta_{\rm so} = 0.01$--0.04
in Cu and Al as listed in Table~1.  Therefore, Eq.~(\ref{eq:VH}) yields
$R_{\rm H}$ of the order of $1${\,}m$\W$, indicating that the spin-current 
induced SHE is observable by using the nonlocal geometry.

\bigskip
\noindent
{\bf Acknowledgement}
\medskip

The authors thank M. Ichimura, H. Imamura, and T. Yamashita for valuable
discussions.  This work is supported by a Grant-in-Aid for Scientific Research
from MEXT and the NAREGI Nanoscience Project.



\end{document}